# In-plane conductivity of a layered large-bipolaron liquid

David Emin

Department of Physics and Astronomy

University of New Mexico

Albuquerque, NM 87131

Distinctive normal-state properties of cuprate superconductors follow from their charge carriers forming a large-bipolaron liquid. The very weak scattering of the liquid's slow moving heavy-massed excitations by acoustic phonons yields a scattering rate that is less than the Debye frequency. The liquid's moderate mobility, $> 1$ cm$^2$/V-sec at 300 K, results from its weak scattering compensating for its large mass. In resolution of a long-standing dilemma, the dc resistivity resulting from scattering by long-wavelength phonons remains nearly proportional to temperature to well below the Debye temperature. Above the Debye frequency the frequency-dependent conductivity is dominated by excitation and photo-ionization of the liquid's self-trapped electronic carriers. Below the Debye frequency the frequency-dependent conductivity is dominated by the liquid's Drude-like collective motion. The "gap" between these two domains sharpens with decreasing temperature as phonon scattering of the liquid's collective excitations diminishes. The high-frequency electronic excitations survive in the superconducting state.

# I. INTRODUCTION

It has been over half a century since it was suggested that mobile local electronic singlet pairs could produce a type of superconductivity akin to the superfluidity found in liquid $^4$He.[1] Furthermore, it has long been realized that electronic carriers can self-trap as singlet pairs to form bipolarons if the atoms surrounding the carriers are sufficiently displaceable.[2,3] In particular, bipolaron formation is stable when the additional energy lowering arising from the self-trapped carriers' pairing overcomes their mutual Coulomb repulsion.

Theoretical investigations indicate that there are two distinct types of polaron and, by analogy, bipolaron.[4-6] A *large* polaron's self-trapped carrier extends over multiple equivalent sites with a spatial extent that varies continuously with the physical parameters. By contrast, a *small* polaron's electronic carrier essentially collapses to a single equivalent site. Large-polaron formation in multi-dimensional electronic systems is associated with the predominance of the Coulomb-interaction-based long-range electron-phonon interaction between the electronic carrier and distant ions. The collapse into a small-polaron can be driven by a sufficiently strong short-range (deformation-potential-like) electron-phonon interaction.

A large polaron moves very slowly but *coherently* in concert with atoms' vibratory movements.[6] The predominance of the long-range electron-phonon interaction then insures that the change of the electronic energy accompanying nearest-neighbor transfer is less than its transfer energy. Moreover, the electronic polarization of a large polaron's self-trapped carrier in response to atoms' vibrations lowers their associated stiffness constant.[6-8] Thus, a large polaron presents a softened region which impedes the transmission of impinging phonons. However, a large-polaron's huge mass enables it to be only weakly scattered by the phonons it encounters.

As a result a large polaron displays a significant room-temperature mobility (e.g. $\mu \geq 1$ cm$^2$/V-s at 300 K) that falls with rising temperature.

By contrast, a small polaron moves *incoherently* by a succession of thermally activated hops.[9-11] The predominance of the short-range electron-phonon interaction then insures that the change of the electronic energy accompanying nearest-neighbor transfer exceeds its transfer energy.[6] The small-polaron mobility is generally very low (e.g. $\mu \leq 1$ cm$^2$/V-s at 300 K) and thermally activated. In particular, a small-polaron's mobility becomes Arrhenius as the temperature is raised above that characterizing the phonons with which the electronic carrier effectively interacts.

Coherent motion of individual quasi-particles is a prerequisite for a superconductor's large-scale coherence. Therefore only large bipolarons provide a suitable basis for bipolarons' superconductivity.[12-14]

A multi-dimensional large-bipolaron can form in a material in which the long-range component of the electron-phonon interaction is associated with a sufficiently large ratio of static to optical dielectric constants $\varepsilon_0/\varepsilon_\infty > 2$.[6,13,15,16] In addition, stabilization of the large bipolaron with respect to two well-separated large polarons requires overcoming the residual repulsion energy $e^2/\varepsilon_0 R_b$, where $R_b$ denotes the radius of the self-trapped pair. The additional bipolaronic binding energy generated by the short-range component of the electron-phonon interaction can provide this stabilization. However, too strong a short-range component of the electron-phonon interaction collapses a large bipolaron into a small bipolaron. The range of strengths of the short-range component of the electron-phonon interaction within which large-bipolaron formation is stabilized expands as the ratio $\varepsilon_0/\varepsilon_\infty$ increases beyond 2. Thus stable large-bipolaron formation is fostered in materials with exceptionally large values of $\varepsilon_0/\varepsilon_\infty$.[6,13,15,16]

Unusually large values of $\varepsilon_0/\varepsilon_\infty$ ($\gg 2$) are measured in layered cuprates.[17-19] These huge values are attributed to very loosely bound ions that reside off the $CuO_2$ planes. By contrast, electronic charge carriers are widely believed to reside within cuprates' $CuO_2$ layers.

A hallmark of a polaron is the optical excitation of its self-trapped carrier.[20] A large-polaron's absorption has three principal components.[6,21] First, optically liberating a large-polaron's self-trapped carrier produces a broad asymmetric optical absorption band. Second, exciting a large-polaron's self-trapped carrier to higher-energy bound states produces relatively narrow absorption bands that reside below the broad ionization absorption. Third, a large-polaron's coherent motion generates a Drude-like absorption whose exceptionally long relaxation time distinguishes it from that of conventional charge carriers. A large-polaron's Drude-like component is suppressed if (1) its coherent motion is precluded by its trapping or (2) its scattering is squelched by its participation in a superconducting condensate. By contrast, a small-polaron's absorption is associated with the transfer of its severely localized self-trapped carrier to a nearby vacant site.[6,11,21-23] Phonon broadening of these optical transitions' initial and final sites yields a wide absorption band whose width increases most markedly as the temperature is raised above the characteristic phonon temperature. The asymmetries of the small-polaron and large-polaron absorptions distinguish between them.[6,11,21] In particular, the small-polaron absorption falls off more slowly below its peak whereas the large-polaron absorption falls off more slowly above its peak. The absorptions for large- or small-bipolarons are qualitatively similar to those of the corresponding polarons.[6,21]

These features are in general accord with measurements on cuprates. For example, the asymmetric photo-induced absorption peaked at ~0.13 eV that is observed in semiconducting $YBa_2Cu_3O_{6.25}$ is consistent with trapped polarons or bipolarons.[24] Furthermore, the observed photo-induced changes in infrared-active vibrations imply localized carrier-induced structural changes.[24] Measurements of the absorption in the normal state of films of superconducting $YBa_2Cu_3O_{7-\delta}$ indicate (1) a similarly broad asymmetric band peaked at ~0.10 eV, (2) a narrower band peaked at ~0.03 eV and (3) a Drude absorption.[25] The Drude absorption disappears as the temperature is lowered below the superconducting transition temperature ~89 K.[25]

In analogy with the superfluidity of liquid $^4$He, large-bipolaron superconductivity is presumed to be a collective property of a fluid ground-state of a liquid comprising large bipolarons. This superconductivity is suppressed if carriers in their ground-state solidify into an ordered arrangement rather than remaining fluid.[26]

The condensation of large bipolarons into a liquid requires an attraction between them. A phonon-mediated attractive interaction results because the carrier-induced lowering of atoms' zero-point vibration energy increases with the carrier density.[27,28] That is, coherence enhances the carrier-induced lowering of phonons' frequencies when their wavelengths exceed the inter-carrier separation $s$. This attractive interaction between large bipolarons behaves roughly as $-\hbar\omega_o(R_b/s)^4$ for $s > 2R_b$, where $\hbar\omega_o$ denotes the characteristic phonon energy. As in the BCS treatment of metals' superconductivity the attraction is proportional to $\hbar\omega_o$.[27-29] There, as here, the ground-state energy is reduced as electronic carriers' coherent response to atomic vibrations lowers their zero-point energy.

There are also repulsive interactions between large bipolarons.[28]  Their mutual long-range Coulomb repulsion $(2e)^2/\varepsilon_0 s$ is reduced by the material's static dielectric constant $\varepsilon_0$.  A short-range repulsion prevents two large-bipolarons merging to form a "quadpolaron."  Since the Pauli principle dictates that a self-trapping potential well's non-degenerate lowest energy level be occupied by no more than one singlet pair, additional carriers must be promoted to excited states. The resulting short-range repulsive interaction between large bipolarons is modeled as $E_b$ $\exp(-s/2R_b)$, where $E_b$ denotes the large-bipolaron binding energy.

With a large enough static dielectric constant $\varepsilon_0 \gg 1$ large-bipolarons' mutual Coulomb repulsions are suppressed sufficiently so that the net two-body interaction potential develops an intermediate-range potential well.[27,28]  The formation of such a potential well enables large-bipolarons' condensation into a liquid at sufficiently low temperatures.  A similar two-body attraction drives $^4$He atoms condensing into a liquid phase.

As the temperature is lowered further carriers ultimately condense into their ground-state and move collectively.  A form of superconductivity akin to the superfluidity of $^4$He can be envisioned if the ground-state remains fluid.  However, this type of superconductivity does not occur if the carriers condense into an ordered arrangement.  In particular, superconductivity is suppressed for carriers that order (locally or globally) in a manner commensurate with the underlying lattice.

Some experimental studies conclude that the charge carriers in the normal state of cuprate superconductors condense into a liquid.[30-33]  Furthermore, cuprates' superconductivity is found to disappear when all carriers condense into an ordered phase.  In particular, the superconductivity of doped $La_2CuO_4$ disappears when the dopants provide one hole for each eight unit cells.[34]  This doping level corresponds to one bipolaron for every $4 \times 4$ superlattice unit of the tetragonal structure: $1/8 = 2/(4 \times 4)$.[6,28]  This suppression of superconductivity is analogous to the suppression of conductivity in $Fe_3O_4$ when polarons condense into an ordered phase as the temperature is lowered below the Verwey transition.[35,36]

The density of carriers in the $CuO_2$ planes of cuprate superconductors is best established in divalently-doped $La_2CuO_4$.  In this case the superconducting regime corresponds to just one bipolaron for between 9 and 25 unit cells.  This carrier density is much less than that in conventional metal-based superconductors, about one electronic carrier per unit cell.  Indeed, no more than moderate carrier densities are possible for large bipolarons since each must be able to shift the equilibrium positions of several of the highly displaceable ions that surround it.

This paper calculates the in-plane conductivity of a layered large-bipolaron liquid generated by its scattering with acoustic phonons.  Section II of this paper addresses the liquid's excitations.  A large bipolarons' huge effective mass insures that the energetic dispersion of the liquid's excitations is much less than the acoustic phonons' Debye energy.  Section III computes the scattering of a planar large-bipolaron liquid by relatively fast-moving acoustic phonons.  In particular, conservation of energy and momentum dictate that slow-moving excitations "reflect" these acoustic phonons.  Distinctively, the rate for acoustic phonons' scattering of the liquid is smaller than the Debye frequency.  Furthermore, this weak scattering compensates for the liquid's huge effective mass to produce a moderate mobility, $> 1$ $cm^2$/V-sec at 300 K.  Strikingly, the scattering rate and the in-plane dc resistivity remain proportional to temperature at temperatures that are well below the Debye temperature.  The frequency-dependent conductivity associated with the liquid's collective motion is primarily manifested at frequencies below the Debye frequency. Above the Debye frequency the frequency-dependent conductivity arises mainly from photo-excitation of large bipolarons' self-trapped electronic carriers.

These unusual features explain long-standing distinctive observations in cuprate superconductors. In particular, cuprate superconductors' resistivity remains proportional to temperature even well below their Debye temperature.[37-42] In addition, the fall-off of cuprate superconductors' frequency-dependent conductivity with diminishing temperature is most pronounced at frequencies below the Debye frequency. Above the characteristic phonon frequencies the frequency-dependent conductivity becomes nearly temperature independent and survives below the superconducting transition temperature.

## II. EXCITATIONS OF A LARGE-BIPOLARON LIQUID

Stable large-bipolaron formation requires an exceptionally large value of the static dielectric constant $\varepsilon_0$ relative to the optical dielectric constant $\varepsilon_\infty$. The static dielectric constant also must be large enough to suppress large-bipolarons' mutual Coulomb repulsion to enable their intermediate-range attraction to drive condensation into a liquid. Therefore this treatment of the conductivity of a large-bipolaron liquid proceeds under the assumption that $\varepsilon_0 >> \varepsilon_\infty > 1$.

An isolated large-bipolaron's transport depends on its phonon scattering. A large bipolaron reduces the stiffness constants that govern the vibrations of the atoms whose displacements bind its self-trapped electronic carrier. Large bipolarons and the solid's phonons thereby scatter one another.  The efficacy of this scattering peaks when the phonons' wavelengths are comparable to a large-bipolaron's spatial extent.[7]

By contrast, phonons' interactions with a large-bipolaron liquid are associated with its excitations.  These excitations describe departures of the liquid from homogeneity.  The strength of this scattering peaks when the phonons' wavelengths are comparable to the wavelengths of the bipolaron-liquid's excitations.

The energy spectrum for excitations of a liquid of large-bipolarons depends on their mutual interactions.  The excitation spectrum at long wavelengths is dominated by the long-range Coulomb repulsions between the liquid's large bipolarons.  At shorter excitation wavelengths the shorter-range interactions between large bipolarons manifest themselves.

In particular, consider a liquid condensate of large-bipolarons that move two-dimensionally within planes that are separated from each other by the distance $b$.  At low enough carrier densities the energy $E(k)$ of an in-plane collective excitation whose wavevector has the magnitude $k$ approaches[43]

$$E(k) = \left[\frac{n_2}{b}\frac{(\hbar k)^2}{m_b}V(k)\right]^{1/2}$$

$$= \sqrt{\frac{2\pi n_2}{m_b}(\hbar k)^2\left\{\frac{(2e)^2}{\varepsilon_0 k}coth\left(\frac{kb}{2}\right)+\left(\frac{2e^2 R_b}{\varepsilon_\infty}\right)\frac{1}{[1+(2kR_b)^2]^{3/2}}\right\}}, \quad (1)$$

where $n_2$ represents the intra-layer large-bipolaron density, $m_b$ depicts the large-bipolaron mass for adiabatic intra-planar movement and $V(k)$ denotes the net Fourier transform for the carriers' mutual interactions. The explicit expression for $V(k)$ derived in Appendix A was introduced to obtain the expression that follows the final equality. For simplicity, the liquid's intermediate-range attractions are ignored. Here $2e$ and $R_b$ denote a large-bipolaron's charge and planar radius, respectively. Physically meaningful values of $k$ are restricted to less than $1/R_b$.

The first contribution within the curly brackets results from the long-range Coulomb repulsions between large bipolarons diminished by the material's static dielectric constant. This contribution dominates as $k \to 0$. The second contribution within the curly brackets depicts short-range intra-layer repulsions between large bipolarons. These short-range interactions increasingly affect the excitations of a large-bipolaron liquid as $k$ is increased from zero.

In the physically significant regime $kR_b << 1$ this excitation energy becomes

$$E(k) \cong \hbar \sqrt{\omega_p^2 \left[ \left( \frac{kb}{2} \right) coth \left( \frac{kb}{2} \right) \right] + \left[ 8\pi \left( \frac{n_2}{m_b} \right) \left( \frac{e^2}{2\varepsilon_\infty R_b} \right) \right] (kR_b)^2}$$

$$\to \hbar \sqrt{\omega_p^2 + \omega_s^2 (kR_b)^2}, \quad (2)$$

where the relation following the arrow pertains when $\varepsilon_0/\varepsilon_\infty$ is so large that the first term within the square-root only manifests itself when $k$ is so small that $kb/2 < 1$. Here the large-bipolaron-liquid's plasma frequency defined by

$$\omega_p \equiv \sqrt{\frac{4\pi(n_2/b)(2e)^2}{\varepsilon_0 m_b}} \quad (3)$$

governs the strength of the Coulomb repulsions and the parameter governing the strength of the short-range repulsions is

$$\omega_s \equiv \sqrt{8\pi\left(\frac{n_2}{m_b}\right)\left(\frac{e^2}{2\varepsilon_\infty R_b}\right)}. \quad (4)$$

The large bipolaron's huge mass, the large static dielectric constant $\varepsilon_0 >> \varepsilon_\infty > 1$ and the relatively small planar carrier density $n_2 << (1/R_b)^2$ team to produce a plasma energy that is even smaller than the Debye energy. For example, the large-bipolaron liquid's plasma energy is no more than that estimated using the effective mass obtained with the adiabatic continuum model. The mass of a planar continuum large bipolaron of thickness $a$ is $m_b \sim e^2/\varepsilon_\infty \omega_o{}^2 a^3$, where $\omega_o$ denotes the Debye frequency with $R_b >> a$ and $b > a$.[15] The resulting plasma energy for a large-bipolaron liquid, $\hbar\omega_o[16\pi(n_2 a^3/b)(\varepsilon_\infty/\varepsilon_0)]^{1/2}$, is usually much smaller than the Debye energy $\hbar\omega_o$. The width of the large-bipolaron excitation spectrum, $\sim E(1/R_b) - E(0)$, is comparably narrow, $\sim \hbar\omega_o[4\pi(n_2 a^3/R_b)]^{1/2} << \hbar\omega_o$.

As illustrated in Fig. 1, $E(k)$ rises from $\hbar\omega_p$ as $k$ is increased from zero. The magnitude of the associated planar group velocity defined by

$$v(k) \equiv \frac{1}{\hbar}\frac{\partial E(k)}{\partial k} \quad (5)$$

is plotted in Fig. 2. For sufficiently small values of $k$, $v(k)$ rises from zero linearly with increasing $k$. For large enough values of $k$ $v(k)$ approaches the $k$-independent constant $\omega_s R_b$. In particular,

$$v(k) = \omega_s R_b \frac{\omega_s k R_b}{\sqrt{\omega_p^2 + \omega_s^2(kR_b)^2}} < \omega_s R_b \sim \omega_o a\sqrt{4\pi(n_2 a R_b)}. \quad (6)$$

As the large-$k$ limit is approached $v(k)$ becomes independent of $b$ since short-range intra-layer repulsions then predominate.

In summary, a large-bipolaron-liquid's excitation spectrum has a width that is less than the Debye energy. Concomitantly the associated excitation group velocities are smaller than the corresponding sound velocity, $< \omega_o a/\pi$.

## III. CONDUCTIVITY OF A LARGE-BIPOLARON LIQUID

A diagonal element of the electrical conductivity tensor for a large-bipolaron liquid's flow parallel to the model's planes is written simply as the product of three terms:

$$\sigma_{xx} \cong \left(\frac{n_2}{b}\right)(2e)\left[\frac{(2e)\tau}{m_b}\right], \quad (7)$$

the density of the liquid's large bipolarons, a bipolaron's charge and the liquid's mobility. The mobility is expressed as the product of an individual bipolaron's charge $2e$ and the time characterizing momentum transfer between the solid's phonons and the liquid's excitations $\tau$ divided by a large-bipolaron's effective mass. In forcing the mobility into this deceptively simple form the number of large-bipolarons in the liquid $N_l$ multiplied by the ratio of the change of the liquid's velocity $\Delta V_l$ produced by a change of its momentum $\Delta P_l$ has been identified with the reciprocal of a large-bipolaron mass: $N_l (\Delta V_l / \Delta P_l) \equiv 1/m_b$. Concomitantly, the distinctive features of the liquid's scattering with phonons are consigned to the calculation of $\tau$.

The scattering between phonons and large-polaronic excitations is unlike that of conventional charge carriers. As illustrated in Fig. 3, an inelastic scattering of a conventional wide-band charge carrier usually involves the absorption or emission of a single phonon of energy $\hbar\omega_q$ and momentum $\boldsymbol{q}$. Then conservation of energy and momentum requires that:

$$E(\boldsymbol{k} \pm \boldsymbol{q}) = E(\boldsymbol{k}) \pm \hbar\omega_{\boldsymbol{q}}. \quad (8)$$

However, this condition generally cannot be fulfilled for large-polaronic excitations because their energetic dispersion is usually much less than the phonon energy.[7] For acoustic phonons these single-phonon processes are precluded when $\left| E(\boldsymbol{k} \pm \boldsymbol{q}) - E(\boldsymbol{k}) \right| \approx \left| \boldsymbol{v}(\boldsymbol{k}) \cdot \hbar\boldsymbol{q} \right|$ is less than $\hbar\omega_{\boldsymbol{q}}$ $= \hbar s \left| \boldsymbol{q} \right|$, where $s$ denotes the sound velocity. Thus single acoustic-phonon scattering processes are suppressed for large-polaronic excitations since their characteristic velocities are less than the sound velocity $\left| \boldsymbol{v}(\boldsymbol{k}) \right| < s$.

Excitation velocities for a large-bipolaron liquid are significantly smaller than the system's sound velocity. Indeed, Eq. (6) shows that the maximum excitation velocity is much less than the sound velocity at the low carrier densities of interest, $n_2 R_b{}^2 << 1$. Scattering of these excitations are then dominated by two-phonon processes in which one phonon is absorbed and another is emitted. Conservation of energy and momentum for this process is described by:

$$E(\boldsymbol{k} \pm \boldsymbol{q} \mp \boldsymbol{q}') = E(\boldsymbol{k}) \pm \hbar\omega_{\boldsymbol{q}} \mp \hbar\omega_{\boldsymbol{q}'}. \quad (9)$$

The weak dispersion of the large-polaronic excitations implies that the energies of the absorbed and emitted phonon must nearly equal one another. In other words, as illustrated in Fig. 4, phonons are essentially "reflected" (elastically scattered) from relatively slow moving polaronic excitations.

Resistance to the liquid's flow occurs as its excitations exchange momentum with phonons. The rate with which acoustic phonons scatter from the liquid's excitations is

$$\frac{1}{\tau(k)} = \frac{1}{(2\pi)^2} \int_0^{q_D} dq\, q \int_0^{2\pi} d\theta \, \frac{1}{[exp(\hbar s q/\kappa T) - 1]} \sigma(k,q) V_x \left[ \frac{|\Delta V_x(\boldsymbol{q})|}{V_x} \right]$$

$$= \frac{1}{(2\pi)^2} \int_0^{q_D} dq\, q \int_0^{2\pi} d\theta \, \frac{1}{[exp(\hbar s q/\kappa T) - 1]} \sigma(k,q) |\Delta V_x(\boldsymbol{q})|, \quad (10)$$

where $q_D \equiv 2\sqrt{\pi}/a$ represents the Debye wave-vector for two-dimensional acoustic phonons. The first factor within the integrand is the Bose factor for these acoustic phonons, $\sigma(k,q)$ denotes the planar "cross-section" for the collision of a phonon of wave-vector of magnitude $q$ with the liquid's excitation of wave-vector of magnitude $k$, $V_x$ represents the liquid's velocity in the $x$-direction and $\left| \Delta V_x(\boldsymbol{q})/V_x \right|$ indicates the scattering-induced fractional change of the liquid's velocity. Furthermore, the change of the $x$-component of the velocity of a large bipolaron produced by an excitation's "reflection" of an acoustic phonon of momentum $\hbar q_x$ is $\Delta V_x(k,\boldsymbol{q}) = 2\hbar q_x/m_b$. With this substitution in Eq. (10) the scattering rate becomes

$$\frac{1}{\tau(k)} = \frac{1}{(2\pi)^2} \int_0^{q_D} dq\, q \int_0^{2\pi} d\theta \, \frac{1}{[exp(\hbar s q/\kappa T) - 1]} \sigma(k,q) \frac{|2\hbar q cos\theta|}{m_b}$$

$$= \frac{8\hbar}{(2\pi)^2 m_b} \int_0^{q_D} dq \, \frac{q^2}{[exp(\hbar s q/\kappa T) - 1]} \sigma(k,q). \quad (11)$$

The characteristic rate for phonons imparting momentum to a large-bipolaron liquid is found by averaging $1/\tau(k)$ over the equilibrium distribution of the liquid's planar excitations $f(k)$:

$$\frac{1}{\tau} \cong \frac{\int_{k_{min}}^{k_{max}} dk\, kf(k)[1/\tau(k)]}{\int_{k_{min}}^{k_{max}} dk\, kf(k)}$$

$$= \frac{2\hbar}{\pi^2 m_b} \frac{\int_{k_{min}}^{k_{max}} dk\, k \left\{ \frac{1}{exp[E(k)/\kappa T]-1} \right\} \int_0^{q_D} dq \left[ \frac{q^2}{exp(\hbar s q/\kappa T)-1} \right] \sigma(k,q)}{\int_{k_{min}}^{k_{max}} dk\, k \left\{ \frac{1}{exp[E(k)/\kappa T]-1} \right\}}, \quad (12)$$

where $k_{max} = 1/d_L$ denotes the reciprocal of the characteristic separation between the liquid's large bipolarons and $k_{min}$ is comparable to the reciprocal of a droplet's characteristic diameter. Here the excitations of the large-bipolaron liquid have been treated as bosons with a vanishing chemical potential since their density is unconstrained.

The cross-section for an individual large polaron's scattering with acoustic phonons rises toward a maximum as $q$ increases from zero.[7] This maximum is achieved when $q$ reaches the reciprocal of the length over which the large polaron's self-trapped charge is redistributed as it softens associated phonons. That is, the maximum of this cross-section occurs when $q$ is comparable to the reciprocal of the large-polaron's diameter. This maximum cross-section governs a large polaron's scattering unless the temperature is so low that even the associated long-wavelength acoustic phonons are frozen out. Furthermore, the maximum cross-section for an individual planar large polaron is simply its diameter $2R_p$.

This paper addresses the scattering of a large-bipolaron liquid's excitations with acoustic phonons at all but extremely low temperatures. The charge redistribution of the liquid is governed by its excitations. The length scales governing these excitations and the acoustic phonons are their respective wavelengths. In this situation the cross-section for two-dimensional acoustic phonons' scattering with the large-bipolaron liquid's planar excitations is represented as the Dirac delta function $\sigma(k,q) \approx \delta(k-q)$.

Upon employing this Dirac delta function Eq. (12) becomes

$$\frac{1}{\tau} = \frac{2\hbar}{\pi^2 m_b} \frac{\int_{k_{min}}^{k_{max}} dk \left\{ \frac{k}{exp[E(k)/\kappa T]-1} \right\} \left[ \frac{k^2}{exp(\hbar sk/\kappa T)-1} \right]}{\int_{k_{min}}^{k_{max}} dk \left\{ \frac{k}{exp[E(k)/\kappa T]-1} \right\}}. \quad (13)$$

Thus the liquid's relaxation rate is proportional to the average of the final squared-bracketed expression in the numerator of Eq. (13) over the equilibrium distribution of the excitations' wave-vectors. The dispersion of the large-bipolaron liquid's excitation energies is weaker than that of the acoustic phonons with which it interacts. In particular, as discussed surrounding Eq. (6), the excitation velocities are much less than the sound velocity. This situation is analogous to that of the large polaron formed by an electronic carrier's interaction with acoustic phonons. Then the velocity of a large-polaron is always much less than the sound velocity.[7]

The strong-scattering relaxation rate for an isolated large-polaron is obtained from Eq. (13) by suppressing the integrations over $k$ and replacing the value of $k$ in the square-bracketed expression by the reciprocal of the large polaron's characteristic length $1/2R_p$.[7] In the high-temperature regime in which the large-polaron's strong-scattering limit prevails, $\kappa T > \hbar s/2R_p$, the relaxation rate reduces to its established value $\sim \kappa T/m_p s R_p$, where $m_p$ denotes the large-polaron's effective mass.[7]

The high-temperature scattering rate for a large-bipolaron liquid, like that for an isolated large polaron, is proportional to the temperature and inversely proportional to the relevant polaronic effective mass. In particular, the high-temperature limit of Eq. (13) becomes

$$\frac{1}{\tau} = \frac{2\kappa T}{\pi^2 s m_b} \frac{\int_{k_{min}}^{k_{max}} dk \left[ \frac{k}{\sqrt{\omega_p^2 + \omega_s^2 (kR_b)^2}} \right] k}{\int_{k_{min}}^{k_{max}} dk \left[ \frac{k}{\sqrt{\omega_p^2 + \omega_s^2 (kR_b)^2}} \right]}, \quad (14)$$

where the final expression for $E(k)$ from Eq. (2) has been utilized. The integrals in Eq. (14) are readily evaluated. However, these integrals become trivial for the parameters that would apply to a large-bipolaron liquid in cuprates: $\varepsilon_0 >> \varepsilon_\infty$ and $R_b \sim b$, the inter-planar separation. This domain is characterized by $\omega_s >> \omega_p$. Then

$$\frac{1}{\tau} \cong \frac{2\kappa T}{\pi^2 s m_b} \frac{\int_{k_{min}}^{k_{max}} dk\, k}{\int_{k_{min}}^{k_{max}} dk} = \frac{\kappa T}{\pi^2 s m_b} (k_{max} + k_{min}). \quad (15)$$

The domain of validity of Eq. (15) is addressed by estimating the correction that emerges as the temperature is lowered. The acoustic phonon's Bose factor then falls from its high-temperature limit, $\kappa T/\hbar sk$ to $(\kappa T/\hbar sk)[1 - (\hbar sk/2\kappa T)]$ as $\hbar sk/\kappa T$ is increased from zero. Upon incorporating this correction factor Eq. (15) becomes

$$\frac{1}{\tau} \cong \frac{2\kappa T}{\pi^2 s m_b} \frac{\int_{k_{min}}^{k_{max}} dk \; k \left[1 - \left(\frac{\hbar sk}{2\kappa T}\right)\right]}{\int_{k_{min}}^{k_{max}} dk}$$

$$= \frac{\kappa T}{\pi^2 s m_b}(k_{max} + k_{min})\left[1 - \frac{\hbar s}{3\kappa T}\left(\frac{k_{max}^3 - k_{min}^3}{k_{max}^2 - k_{min}^2}\right)\right]. \quad (16)$$

Thus significant corrections to Eq. (15) only occur when the temperature is lowered below about $\hbar sk_{max}/3\kappa$. Using the definition of the Debye temperature, $T_D \equiv \hbar sq_D/\kappa$, and the definitions of $q_D$ and $k_{max}$ found below Eqs. (10) and (12), it becomes apparent that $\hbar sk_{max}/3\kappa = T_D/3q_D d_L \sim T_D[(1/6\sqrt{\pi})(a/d_L)] << T_D$, since $a << d_L$.

The corresponding planar resistivity is estimated by incorporating Eq. (15) into the reciprocal of the electrical conductivity of Eq. (7) and taking $k_{max} >> k_{min}$:

$$\rho_{xx} = \frac{1}{\left(\frac{n_2}{b}\right)(2e)^2}\frac{m_b}{\tau} = \frac{b}{n_2(2e)^2}\frac{\kappa T}{\pi^2 s}(k_{max} + k_{min}) \to \frac{b}{n_2(2e)^2}\frac{\kappa T}{\pi^2 s d_L}. \quad (17)$$

The final expression on the r.h.s. of Eq. (17) pertains for a global liquid, $k_{max} = 1/d_L$ and $k_{min} = 0$. Then the in-plane resistivity for the large-bipolaron liquid is proportional to both the temperature $T$ and the inter-layer separation $b$ while being inversely proportional to the sound velocity $s$ and the square root of the planar density of the liquid's large bipolarons $n_2$ since $d_L \cong n_2^{-1/2}$. These parameters can be readily estimated to yield a planar resistivity of about 100-1000 $\mu\Omega$-cm at 300 K. These estimates are comparable to values reported for the normal-state in-plane resistivity of cuprate superconductors, e.g. 400 $\mu\Omega$-cm at 300 K in $Sr_{0.2}La_{1.8}CuO_4$.[37-42]

The in-plane frequency-dependent conductivity of a large-bipolaron liquid $\sigma_{xx}(\omega,T)$ is composed of two principal components. At low-frequencies the conductivity is dominated by the Drude conductivity associated with the weak scattering (long scattering time, $\omega_o\tau >> 1$) of the heavy-massed large-bipolaron excitations. At relatively high frequencies the conductivity is dominated by excitations of the self-trapped electronic carriers that free them from their self-trapping potential wells.

The weak-scattering of the excitations of a large-bipolaron liquid of Eq. (15) insures that the Drude contribution to $\sigma_{xx}(\omega,T)$,

$$\sigma_{xx}(\omega,T)|_{Drude} = \frac{\sigma_{xx}(0,T)}{1+(\omega\tau)^2}, \quad (18)$$

falls off at frequencies below the Debye frequency $\omega_o$. Since the scattering time of Eq. (15) is inversely proportional to temperature, the frequency characterizing the fall-off of the Drude contribution, proportional to $1/\tau$, falls as the temperature is lowered. These behaviors are illustrated in Fig. 5.

A peaked high-frequency temperature-independent contribution to $\sigma_{xx}(\omega,T)$ is produced by the in-plane photo-ionization of self-trapped carriers of electronic mass $m$ from planar large-bipolarons,[6,21]

$$\sigma_{xx}(\omega)|_{ionizing} = \frac{8\pi^2(2n_2/b)e^2}{m\omega}\left[\frac{\left(\frac{\hbar\omega}{E_p}-3\right)}{\left[\left(\frac{\hbar\omega}{E_p}-2\right)^3\right]}\right], \quad (19)$$

which occurs above the photo-ionization threshold, here $\hbar\omega > 3E_p$. As described by Eq. (19), for the regime of concern here, $\varepsilon_0 >> \varepsilon_\infty$, the contribution to $\sigma_{xx}(\omega)$ from photo-ionization of planar large-bipolarons approaches that for photo-ionizing twice as many planar large polarons of binding energy $E_p \cong e^2/4\varepsilon_\infty R_p$ whose self-trapped electronic carriers have threshold ionization energy $3E_p$ [c.f. Eq.(11) of Ref. (21)].[21] The threshold ionization energy is just the binding energy of the self-trapped electronic carriers. Self-trapping requires that this energy be above the characteristic phonon energy, $3E_p > \hbar\omega_o \equiv \kappa\theta$. These behaviors are illustrated in Fig. 6.

Estimates of the large-bipolaron effective mass $m_b$ and the electronic effective mass $m$ can be obtained from the contributions to the frequency-dependent conductivity described by Eqs. (18) and (19), respectively. Furthermore, integrals of each of these two contributions over frequency $\omega$, $[(n_2/b)(2e)^2(\pi/2)]/m_b$ and $[(2n_2/b)e^2(4\pi^2/3)]/m$, also provide estimates of $m_b$ and $m$.

Finally it should be noted that additional contributions to the frequency-dependent conductivity occur below the threshold for ionization of the self-trapped electronic carriers. These contributions arise from electronic carriers moving in concert with vibrations of the atoms responsible for the self-trapping potential well and from photo-induced promotion of self-trapped carriers to excited bound states within the self-trapping potential well.[6,21]

## IV. DISCUSSION

This calculation of the dc in-plane conductivity presumes that electronic charge carriers self-trap. Self-trapped electronic carriers move very slowly in concert with atomic movements. To assess the relevance of this work to cuprate superconductors it is useful to summarize evidence of their electronic charge carriers undergoing such localization. First, distinctive absorption bands are produced by exciting trapped electronic carriers from and within the potential wells within which they are confined.[6,21] Such absorptions are observed upon introducing electronic charge carriers to cuprate insulators.[24] Similar absorptions are also observed in the normal and superconducting states of the corresponding cuprate superconductors.[25] Second, the polarization of localized electrons in response to surrounding atoms' vibrations can generate localized vibration modes.[8] These extrinsic vibration modes only occur with wavelengths that are less than the localized-state radii. Such "ghost modes" are observed in cuprates upon introducing electronic charge carriers.[44]  Third, localized states associated with impurities produce a photoemission which is distinctive in that its intensity increases in proportion with the density of impurities.  The carrier-induced photoemission of cuprate superconductors is like that from impurity states' localized carriers.[45,46]  Fourth, localized electrons generally decrease the lifetimes of impinging positrons by providing centers for their annihilation.  The lifetimes of positrons that impinge on cuprates decrease upon introducing electronic charge carriers.[47]  Moreover, these positron lifetimes revert toward those of undoped materials as cuprates are cooled into their superconducting state.  The non-uniformity that characterizes a collection of self-trapped carriers in their normal state is presumably suppressed in their superconducting condensate.  None of these cuprate phenomena are observed in conventional superconductors.

The notion of a large-bipolaron liquid may also provide a basis for understanding the superconductivity of doped ionic semiconductors other than the cuprates.  The carrier density must then be low enough so that competition to displace ions surrounding different large bipolarons does not preclude their formation.  Moreover, large-bipolaron formation requires that the semiconductor contains sufficient displaceable ions so that its static dielectric constant greatly exceeds its optical dielectric constant, $\varepsilon_0 >> 2\varepsilon_\infty$.[6,13,15,16]  Furthermore, systems of reduced electronic dimensionality fosters large-bipolarons' formation by impeding their collapse into small bipolarons.[6,16]  The low-temperature superconductivities ( < 0.3 K) of reduced and Nb-doped $SrTiO_3$ as well as those at $LaAlO_3/SrTiO_3$ and $LaTiO_3/SrTiO_3$ interfaces satisfy these criteria.[48-51]

## V. SUMMARY

Moderate densities of electronic charge carriers in ionic solids having exceptionally large static dielectric constants, $\varepsilon_0 >> 2\varepsilon_\infty$, can self-trap as singlet pairs that extend over multiple sites thereby forming large bipolarons.  The electronic polarizabilities of the self-trapped electronic carriers produce a phonon-assisted attraction between large bipolarons that can drive their condensation into a liquid.  Scattering of the large-bipolaron liquid's excitations by acoustic phonons generates electrical resistance.  The very weak dispersion of the liquid's excitations results in unconventional weak scattering.  Distinctively, the scattering rate is generally less than the Debye frequency and rises nearly in proportion to the temperature even at temperatures well below the Debye temperature.  This very weak scattering compensates for the hefty large-bipolaron mass to produce a moderate mobility, e.g. $> 1$ cm$^2$/V-sec at 300 K.

Distinguishing features of the in-plane conductivity of a layered large-bipolaron liquid agree with those observed in cuprate superconductors.  The dc resistivity is proportional to temperature even at temperatures well below the Debye temperature.  The very weak temperature-dependent scattering of the large-bipolaron liquid insures that the decrease of the conductivity with increasing applied frequency and decreasing temperature is most pronounced below the Debye frequency.  By contrast, the conductivity at frequencies above the Debye frequency arises from the nearly temperature-independent photo-excitation of the large-bipolaron liquid's self-trapped electronic carriers.

## APPENDIX A: INTERACTIONS BETWEEN LARGE BIPOLARONS

Consider mutually interacting carriers that move within parallel planes whose adjacent planes are separated by the distance $b$. The interaction between two large bipolarons is taken to be the sum of their mutual long-range Coulomb repulsion and their short-range repulsion. The short-range repulsion occurs because merger of two singlet bipolarons is energetically unfavorable. In particular, the Pauli principle requires that two of the merged bipolarons' four electronic carriers be promoted into high-lying orbitals. Large bipolarons' short-range mutual repulsion is presumed to only occur between large bipolarons within the same plane.

The sum of these two interactions is modelled as:

$$I(s,n) = \frac{(2e)^2}{\varepsilon_0 \sqrt{s^2 + (nb)^2}} + \delta_{n,0}\left(\frac{e^2}{2\varepsilon_\infty R_b}\right) e^{-s/2R_b}, \quad (A1)$$

where $\varepsilon_0$ and $\varepsilon_\infty$ respectively depict the static and optical dielectric constants   Here $s$ represents the intra-planar separation between large-bipolarons and $n$ is the number of planes separating them. A large bipolaron's binding energy, radius and charge are written as $E_b$, $R_b$, and $2e$, respectively. In the limit that $\varepsilon_0/\varepsilon_\infty \to \infty$, $E_b \to e^2/2\varepsilon_\infty R_b$.

The Fourier transform of the cylindrically symmetric inter-particle interaction function $I(s,n)$ is

$$V(k) = b \sum_{n=-\infty}^{\infty} \int_0^{\infty} ds \, sI(s,n) \int_0^{2\pi} d\theta e^{iks\cos\theta} = 2\pi b \sum_{n=-\infty}^{\infty} \int_0^{\infty} ds \, sI(s,n)J_0(ks), \quad (A2)$$

where polar coordinates are employed in the integrations and $J_0(ks)$ represents the zeroth-order Bessel function. Inserting the interaction function into the expression for its Fourier transform and then changing variables yields

$$V(k) = \frac{2\pi b}{k^2} \sum_{n=-\infty}^{\infty} \int_0^{\infty} du \, uV(u/k, n)J_0(u)$$

$$= \frac{2\pi b(2e)^2}{\varepsilon_0 k} \sum_{n=-\infty}^{\infty} \int_0^{\infty} du \, u \frac{J_0(u)}{\sqrt{u^2 + (nkb)^2}}$$

$$+ \frac{2\pi b e^2}{2\varepsilon_\infty R_b k^2} \int_0^{\infty} du \, ue^{-u/2kR_b}J_0(u). \quad (A3)$$

These two integrals have been evaluated [6.554.1 of I. S. Gradsheteyn and I. M. Ryzhik and 312 18b) of W. Gröbner and N. Hofreiter].[52,53] Incorporating these results and evaluating the summation over the planar index $n$ yields:

$$V(k) = 2\pi b \left\{ \frac{(2e)^2}{\varepsilon_0 k} \sum_{n=-\infty}^{\infty} e^{-|nkb|} + \left(\frac{2e^2 R_b}{\varepsilon_\infty}\right) \frac{1}{[1 + (2kR_b)^2]^{3/2}} \right\}$$

$$= 2\pi b \left\{ \frac{(2e)^2}{\varepsilon_0 k} \coth\left(\frac{kb}{2}\right) + \left(\frac{2e^2 R_b}{\varepsilon_\infty}\right) \frac{1}{[1 + (2kR_b)^2]^{3/2}} \right\}. \quad (A4)$$

Inter-layer interactions between large-bipolarons are suppressed in the limit $kb \to \infty$. Then $\coth(kb/2)$ is replaced by unity within the curly brackets so that the Coulomb repulsion term of the interaction function $V(k)$ becomes proportional to $1/k$. This behavior characterizes strictly two-dimensional Coulomb interactions. By contrast, for a three-dimensional system ($kb$ finite) the Coulomb repulsion term of the interaction becomes proportional to $1/k^2$ as $k \to 0$.

# Figure captions

Fig. 1. The excitation energy $E(k)$ in units of $\hbar\omega_p$ is plotted against $kR_b$ for $b/R_b = 2$ and $\varepsilon_0/\varepsilon_\infty = 35$.

Fig. 2. The group velocity $v(k)$ in units of $\omega_p R_b$ is plotted against $kR_b$ for $b/R_b = 2$ and $\varepsilon_0/\varepsilon_\infty = 35$.

Fig. 3. The strong scattering of a conventional electronic carrier (solid arrow) with a phonon (dashed arrow) is schematically illustrated.

Fig. 4. The weak scattering of a large-polaronic excitation (solid arrow) with phonons (dashed arrows) is schematically illustrated.

Fig. 5 The Drude contribution to the conductivity of a large-bipolaron liquid normalized to its dc value evaluated at the phonon temperature $\theta$ is plotted versus the applied frequency $\omega$ in units of the characteristic phonon frequency $\omega_o$ at three progressively lower temperatures in curves $a$, $b$ and $c$. This illustrative plot presumes $m_b = 10$ m, $s = 5 \times 10^3$ m/s, $d_L = 1$ nm and $\theta/T = 1$, 2 and 3.

Fig. 6 The contribution to the conductivity of a large-bipolaron liquid normalized to its dc value evaluated at the phonon temperature $\theta$ produced by photo-ionizing the liquid's self-trapped carriers is plotted versus the applied frequency $\omega$ in units of the characteristic phonon frequency $\omega_o$. This illustrative plot presumes $s = 5 \times 10^3$ m/s, $d_L = 1$ nm and $E_p/\hbar\omega_o = 2$.

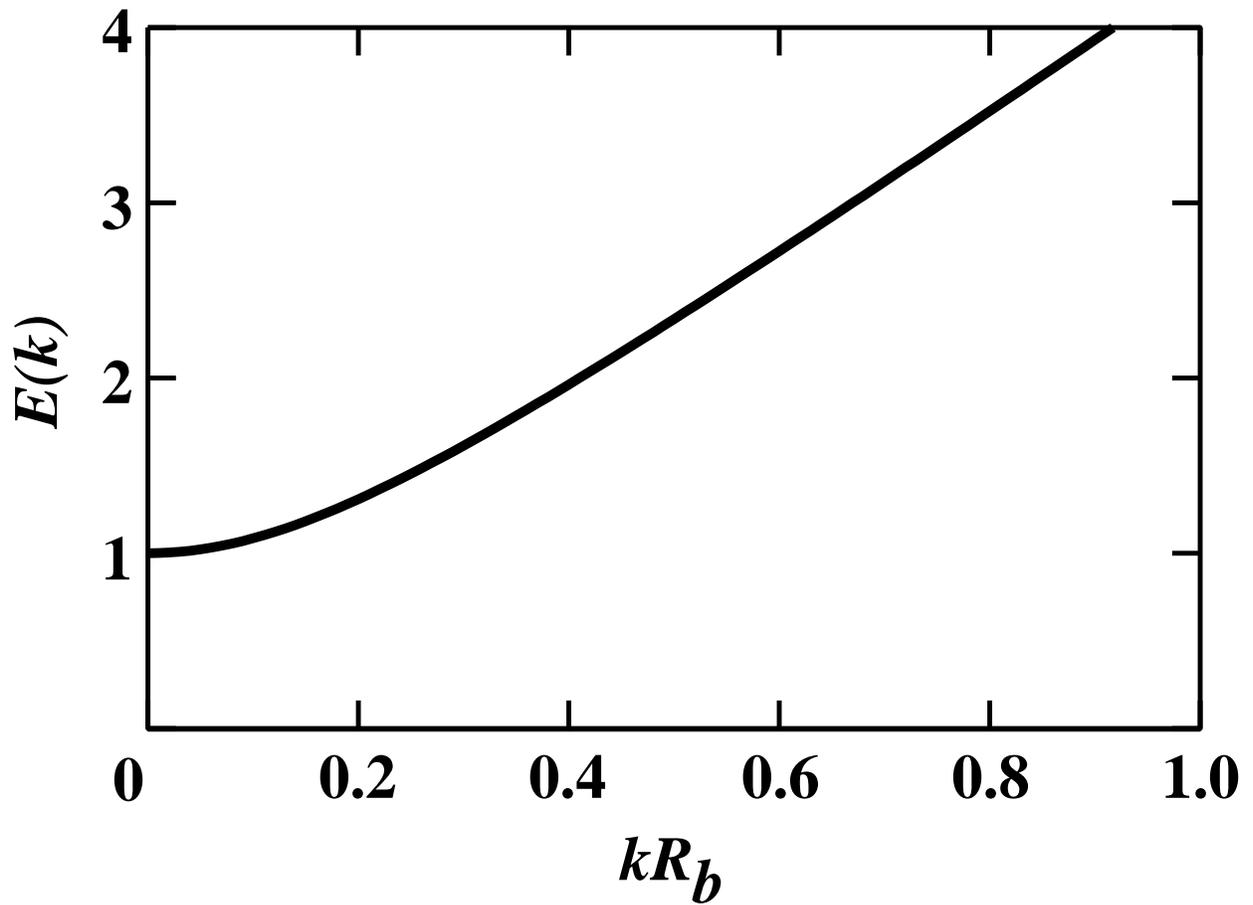

Fig. 1. The excitation energy $E(k)$ in units of $\hbar\omega_p$ is plotted against $kR_b$ for $b/R_b = 2$ and $\varepsilon_0/\varepsilon_\infty = 35$.

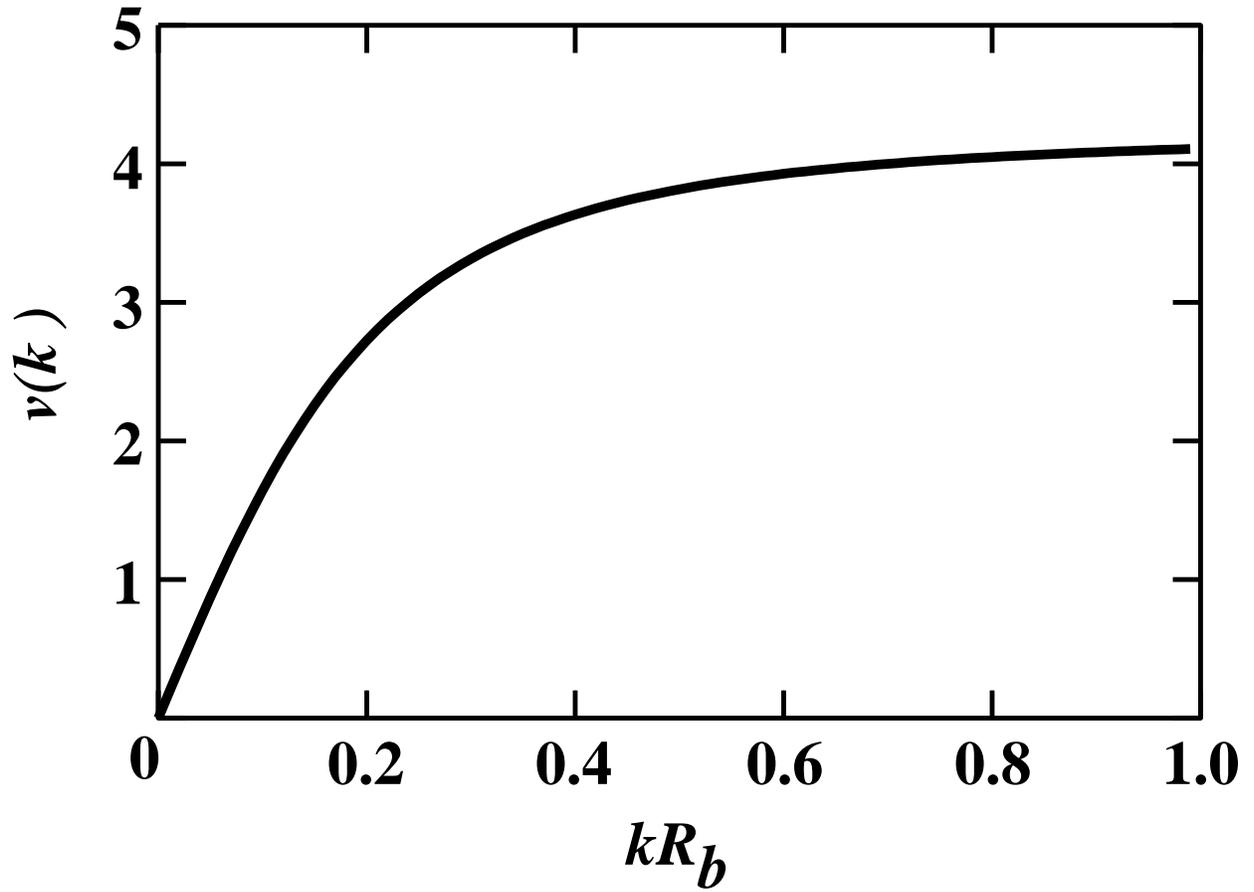

Fig. 2 The group velocity $v(k)$ in units of $\omega_p R_b$ is plotted against $kR_b$ for $b/R_b = 2$ and $\varepsilon_0/\varepsilon_\infty = 35$.

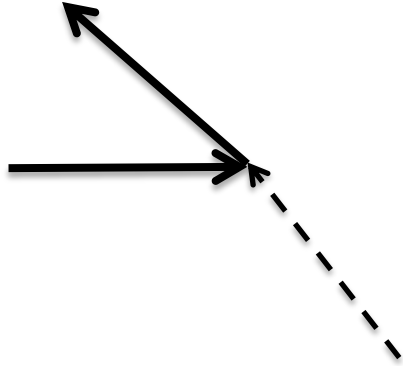

Fig. 3. The strong scattering of a conventional electronic carrier (solid arrow) with a phonon (dashed arrow) is schematically illustrated.

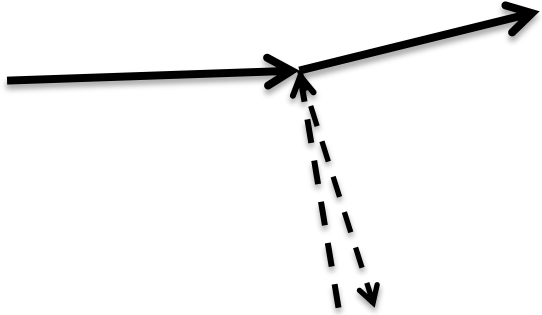

Fig. 4 The weak scattering of a large-polaronic excitation (solid arrow) with phonons (dashed arrows) is schematically illustrated.

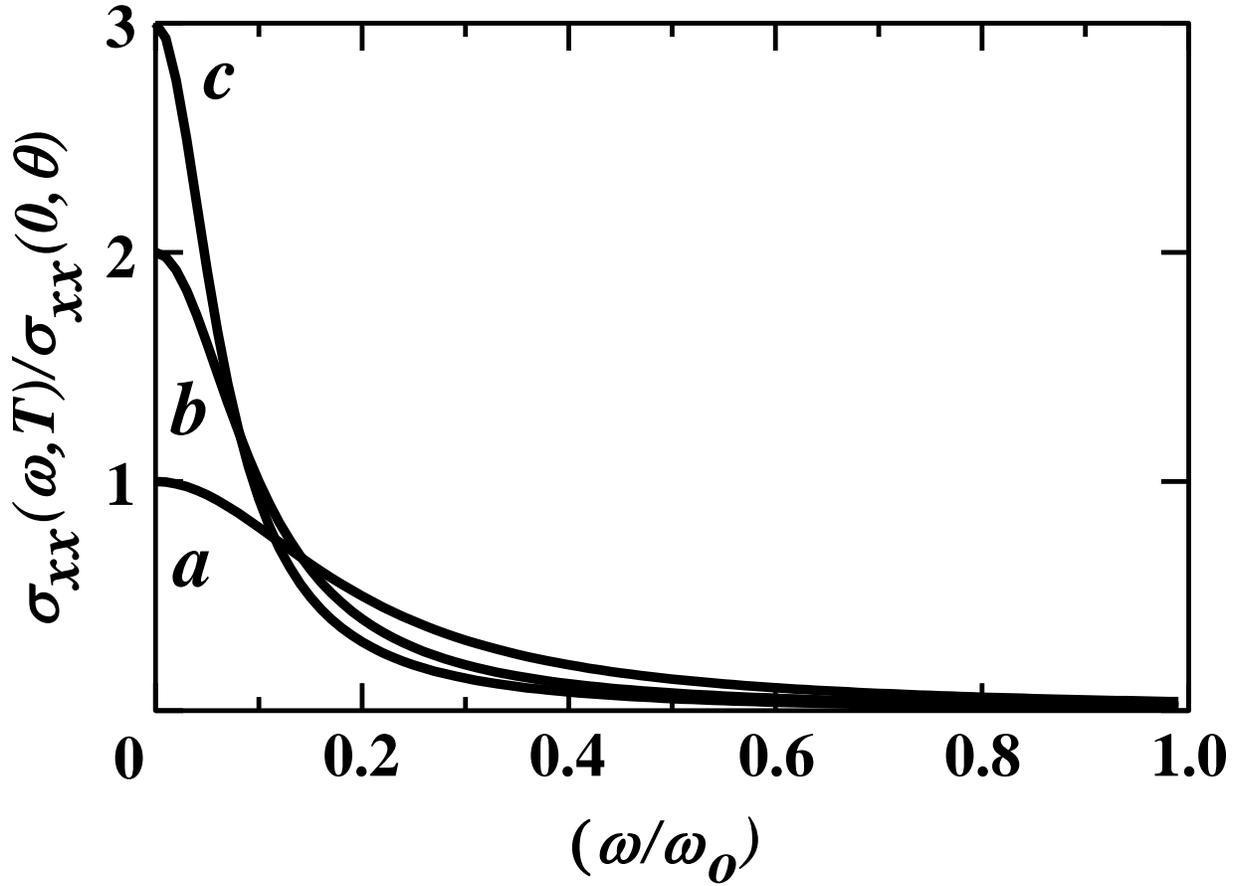

Fig. 5 The Drude contribution to the conductivity of a large-bipolaron liquid normalized to its dc value evaluated at the phonon temperature $\theta$ is plotted versus the applied frequency $\omega$ in units of the characteristic phonon frequency $\omega_o$ at three progressively lower temperatures in curves $a$, $b$ and $c$. This illustrative plot presumes $m_b = 10$ m, $s = 5 \times 10^3$ m/s, $d_L = 1$ nm and $\theta/T = 1$, 2 and 3.

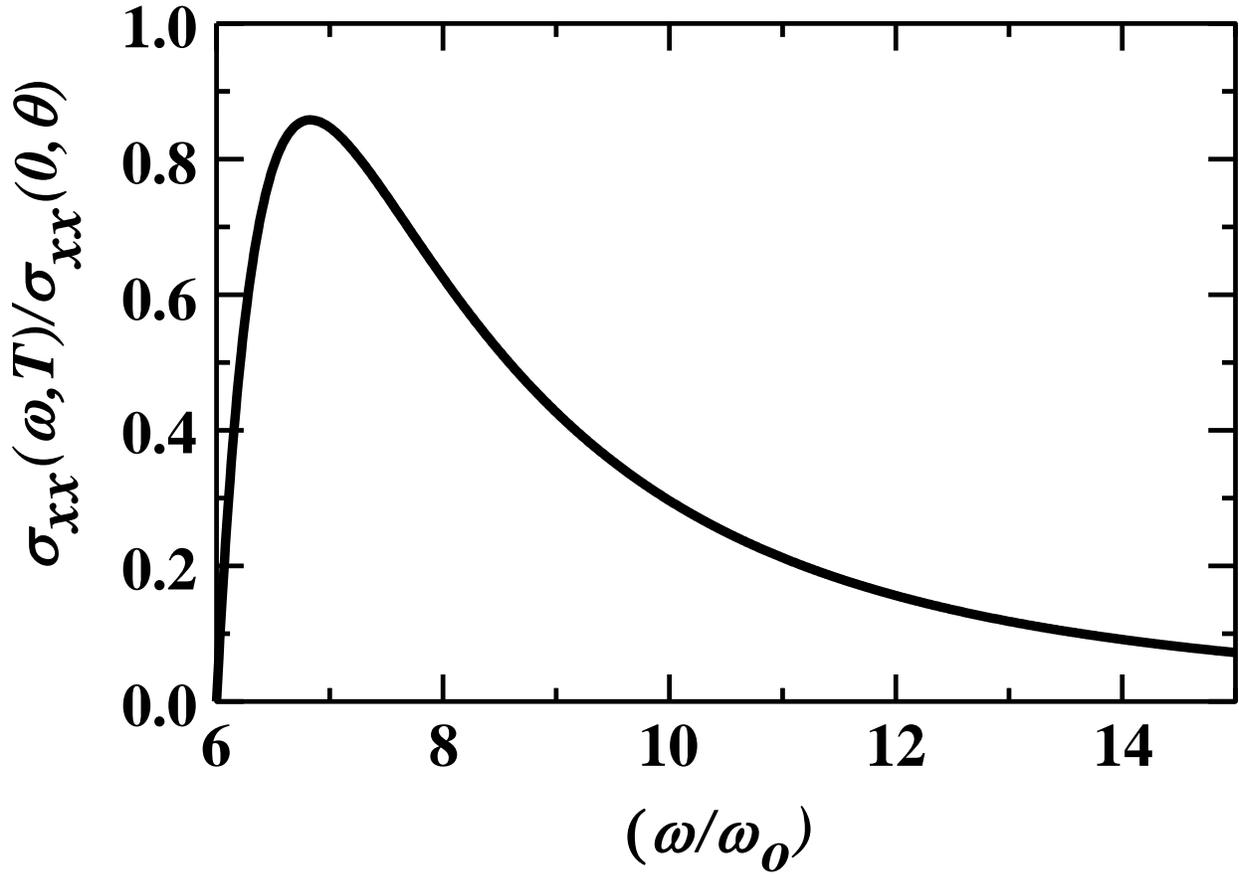

Fig. 6 The contribution to the conductivity of a large-bipolaron liquid normalized to its dc value evaluated at the phonon temperature $\theta$ produced by photo-ionizing the liquid's self-trapped carriers is plotted versus the applied frequency $\omega$ in units of the characteristic phonon frequency $\omega_o$. This illustrative plot presumes $s = 5 \times 10^3$ m/s, $d_L = 1$ nm and $E_p/\hbar\omega_o = 2$.